\documentclass[11pt]{article}

\usepackage{amssymb,amsmath,amsfonts,amsthm,lscape}

 \usepackage{graphicx}

\def\be{\begin{equation}}
\def\ee{\end{equation}}
\def\bea{\begin{eqnarray}}
\def\eea{\end{eqnarray}}

\newcommand{\bq}{\mathbf{q}}
\newcommand{\bp}{\mathbf{p}}

\newcommand{\hbq}{\hat{\mathbf{q}}}
\newcommand{\hbp}{\hat{\mathbf{p}}}
\newcommand{\hH}{\hat{\cal{H}}}
\newcommand{\hC}{\hat{C}}

\newcommand\Om\Omega
\def\NN{\mathbb N}
\def\bb{\beta}

\def\omm{\Omega}

\newcommand{\cH}{{\cal{H}}}

\newcommand{\hq}{\hat{q}}
\newcommand{\hp}{\hat{p}}

\newcommand{\la}{\lambda}

\newcommand{\te}{\theta}
\newcommand{\dd}{{\rm d}}
\newcommand{\kk}{k}

 \def\RR{\mathbb{R}}

 \def\1{\'{\i}}
 \newcommand{\cM}{{\mathcal M}}
  \newcommand{\cR}{{\cal R}}

  \newcommand{\aaa}{\mu}

\def\rmi{{\rm i}}
\newcommand{\pd}{\partial}

\newcommand{\om}{\omega}
\newcommand{\De}{\Delta}

 \def\hamG{\hat{\cal H}_{{\rm c},\la}}
  \def\hamGb{\hat{\cal H}_{{\rm c},\eta}}

\newcommand{\LB}{{\rm c}}

\newcommand{\hI}{\hat{I}}

\begin{document}

\thispagestyle{empty}

\

 \vskip1cm

\noindent {\Large{\bf {Exactly solvable deformations of the oscillator and Coulomb\\[6pt] systems and their generalization
}}}

\medskip
\medskip
\medskip
\medskip

\begin{center}
{\sc \'Angel Ballesteros$^a$, Alberto Enciso$^b$, Francisco J. Herranz$^{a,}$\footnote{
 Based on the contribution presented at ``The 30th International Colloquium on Group Theoretical Methods in Physics",
July 14--18, 2014, 
 Ghent, Belgium. To appear in {\em Journal of Physics: Conference Series}.}
 \\[4pt] Orlando Ragnisco$^{c}$ and Danilo Riglioni$^{d}$}
 \end{center}

\medskip
\medskip

\noindent
{$^a$ Departamento de F{\'{\i}}sica, Universidad de Burgos, E-09001 Burgos, Spain}

\medskip

\noindent{$^b$ Instituto de Ciencias Matem\'aticas,  CSIC, Nicol\'as Cabrera 13-15, E-28049 Madrid,
Spain}

\medskip

\noindent{$^c$ Dipartimento di   Matematica e Fisica, Universit\`a di Roma Tre and Istituto Nazionale di Fisica Nucleare sezione di Roma Tre, Via Vasca Navale 84, I-00146 Roma, Italy}

\medskip

\noindent{$^d$ Centre de Recherches Math\'ematiques, Universit\'e de Montr\'eal, H3T 1J4 2920 Chemin de la tour, Montreal, Canada}

\medskip

\noindent{E-mail:\quad {\tt angelb@ubu.es, aenciso@icmat.es, fjherranz@ubu.es, \\[2pt]  ragnisco@fis.uniroma3.it,
 riglioni@crm.umontreal.ca}}

\medskip

\medskip

\begin{abstract}
\noindent
We present  two maximally superintegrable Hamiltonian systems ${\cal H}_\la$ and ${\cal H}_\eta$ that are defined, respectively, on an $N$-dimensional spherically symmetric generalization of the Darboux surface of type III and on an $N$-dimensional Taub--NUT space. Afterwards, we show that the quantization of ${\cal H}_\la$ and ${\cal H}_\eta$ leads, respectively, to exactly solvable deformations (with parameters $\la$ and $\eta$) of the two basic quantum mechanical systems: the harmonic oscillator and   the Coulomb problem. 
In both cases the quantization is performed in such a way that the maximal superintegrability of the classical  Hamiltonian is fully preserved.  In particular, we   prove that this strong condition is fulfilled by  applying the so-called conformal Laplace--Beltrami quantization prescription, where the {conformal Laplacian} operator contains the usual Laplace--Beltrami operator on the underlying manifold plus a term proportional to its scalar curvature (which in both cases has non-constant value). In this way, the eigenvalue problems for the quantum counterparts of ${\cal H}_\la$ and ${\cal H}_\eta$ can be rigorously solved, and it is found that their discrete spectrum is just a smooth deformation (in terms of the parameters $\la$ and $\eta$) of the oscillator and Coulomb spectrum, respectively. Moreover, it turns out that the maximal degeneracy of both systems is   preserved under deformation. Finally, new  further multiparametric generalizations of both systems that preserve their superintegrability are envisaged.

\end{abstract}

\newpage


\section{Introduction}

It is well known that if we consider a natural classical Hamiltonian system on the $N$-dimensional ($N$D) Euclidean space
\be
{\cal H}={\cal T}(\bp)+{\cal U}(\bq),
\label{uno}
\ee
the harmonic oscillator potential ${\cal U}(\bq)=\om^2 \bq^2$ and the Coulomb potential ${\cal U}(\bq)=-\kk/ |\bq|$ define two {\em maximally superintegrable} (MS) systems (in the Liouville sense), since both systems are endowed with $(2N-1)$ functionally independent and globally defined integrals of the motion. In the first case such integrals are provided by the components of the Demkov--Fradkin tensor~\cite{Demkov, Fradkin}, and in the second one by the angular momenta together with the $N$ components of the Runge--Lenz vector (see e.g.~\cite{commun} and references therein). At the classical dynamical level, the footprint of superintegrability consists in the fact that all bounded trajectories of these two systems are closed ones, a fact which is diretly related with Bertrand's theorem~\cite{Bertrand2}. Moreover, when the quantization of these systems is performed it is found that such superintegrability implies that their spectrum exhibits maximal degeneracy due to a superabundance of quantum integrals of the motion.

In this paper we review two spherically symmetric deformations of the oscillator and Coulomb systems that define two new MS systems~\cite{PD, sigma72011}. As a consequence, their quantization~\cite{darbouxiii, annals2011,annals2014} is shown to present maximal degeneracy in the spectra. At a first sight, the existence of such deformations could seem impossible since   the only spherically symmetric potentials on the Euclidean space that are MS are just the oscillator and the Coulomb ones. Therefore, the addition of any radial perturbation on these systems leads to   superintegrability breaking and thus to a lack of maximal degeneracy in the spectra, a fact that is very well known in quantum perturbation theory. However, as we shall see, such superintegrable perturbations can be obtained if {\em both the potential and the kinetic energy} are simultaneously deformed in a very precise way. Explicitly, the Hamiltonian~\eqref{uno} will be smoothly deformed into
\be
{\cal H}_{\aaa}(\bq,\bp)={\cal T}_{\aaa}(\bq,\bp)+{\cal U}_{\aaa}(\bq),
 \label{dos}
\ee
where $\aaa$ can be regarded as a (generic)  {\em deformation parameter} in such a manner that  we will be no longer working on the {\em flat} Euclidean space, but on a suitable curved space with metric and kinetic energy given by
$$
{\rm d}s_\aaa^2=\sum_{i,j=1}^N g_{ij}(\bq){\rm d}q_i{\rm d}q_j,\quad\  {\cal T}_\aaa(\bq,\bp)=\frac12\sum_{i,j=1}^N g^{ij}(\bq) {p_ip_j} .
$$
 This fact will provide additional interesting geometric features to the systems we will deal with. In particular, we will see that the curved/deformed generalization of the Demkov--Fradkin tensor and of the  Runge--Lenz vector do exist, and will be the essential tool to prove the MS property of the deformed systems. 

We recall that the quantization problem on curved spaces is clearly a non-trivial one, since the kinetic energy term ${\cal T}_{\aaa}(\bq,\bp)$ is a function of both positions and momenta that creates severe ordering ambiguities. Nevertheless, we shall explicitly show that   a quantization of $\cH_\aaa$ (\ref{dos})  that preserves  the   MS property is achieved   through the {\em conformal Laplacian quantization} (see~\cite{annals2011,annals2014, Wa84, Liu, Landsman, MRS} and references therein):
\be
\hat{\mathcal{H}}_{\rm c,\aaa}=\hat{\mathcal T}_{\rm c,\aaa} +{\mathcal U}_\aaa =-\frac{\hbar^2}2 \Delta_{\rm c,\aaa}+{\mathcal U}_\aaa= -\frac{\hbar^2}2 \left(\Delta_{\rm LB,\aaa} -  \frac{  (N-2)}{4(N-1)} \,R_\aaa
\right)+\mathcal U_\aaa\, ,
\label{conf}
\ee
where $R_\aaa$ is the scalar curvature on the underlying $N$D curved manifold  ${\cal M}_\aaa$,   
the operator $ \Delta_{\rm c,\aaa}$ is the conformal Laplacian~\cite{Baer} and  $\Delta_{\rm LB,\aaa}$ is the   usual
 Laplace--Beltrami  operator  on ${\cal M}_\aaa$,  {\em i.e.},
$$
 \Delta_{\rm LB,\aaa}=\sum_{i,j=1}^N \frac 1{\sqrt{g}}\partial_i\sqrt{g} g^{ij}\partial_j \, ,
$$
where $g^{ij}$ is the inverse of the metric tensor $g_{ij}$ and $g$ is the corresponding determinant. 
 The limit $\aaa\to 0$ gives rise to the  quantization of the flat Hamiltonian (\ref{uno})  with $\dd s^2= \dd \bq^2$  since
 $$
    \lim_{\aaa\to 0}  R_\aaa=0,\qquad   \lim_{\aaa\to 0} \Delta_{\rm c,\aaa}= \lim_{\aaa\to 0}\Delta_{\rm LB,\aaa}=\Delta=\nabla^2,
\qquad   \lim_{\aaa\to 0} \hat{\mathcal{H}}_{\rm c,\aaa}=-\frac{\hbar^2}2\nabla^2+ {\mathcal U} \, .
 $$
 We also recall that the quantization  (\ref{conf})  can be related through a similarity transformation to the Hamiltonian obtained by means of the so-called {\em direct Schr\"odinger quantization} prescription on conformally flat spaces~\cite{darbouxiii,annals2011} \begin{equation}
\label{direct}
\hat{\mathcal{H}}_\aaa=\hat{\mathcal T}_{ \aaa} +{\mathcal U}_\aaa   = -\frac{\hbar^2}{2 f_\aaa(r)^2} \,\Delta  +\mathcal U_\aaa\,   , \nonumber
\end{equation}
where $ f_\aaa(r)= f_\aaa(|\bq|)$ is the conformal factor of the metric on   ${\cal M}_\aaa$ written as $\dd s_\aaa^2= f_\aaa(r)^2 \dd\bq^2$.
In this case, the scalar curvature reads
\be
R_\mu=-(N-1)\left( \frac{    (N-4)f_\aaa'(r)^2+  f_\aaa(r)  \left(    2f_\aaa''(r)+2(N-1)r^{-1}f_\aaa'(r)  \right)}   {f_\aaa(r)^4  } \right) .
\label{aac}
\ee

  In the next  two sections,  we review the  exactly solvable deformations of the $N$D  isotropic oscillator $\hat{\mathcal{H}}_{\rm c,\lambda}$~\cite{annals2011} and the Coulomb system $\hat{\mathcal{H}}_{\rm c,\eta}$~\cite{annals2014}, correspondingly.  New results are sketched in the last section by presenting the only possible multiparametric spherically symmetric generalizations of the above systems which are MS with {\em quadratic} integrals of motion, that is, the most generic deformations that can be endowed, respectively, with a  generalized Demkov--Fradkin tensor and with a Runge--Lenz $N$-vector.


\section{An exactly solvable deformation of the oscillator system}

 The $N$D classical Hamiltonian system given by
\be
{\cal H}_\la(\bq,\bp)={\cal T}_\la(\bq,\bp)+{\cal U}_\la(\bq)=
\frac{\bp^2}{2(1+\la \bq^2)}+\frac{ \om^2 \bq^2}{2(1+\la \bq^2)}  ,
 \label{ac}
\ee
where $\la$ and $\om$ are real parameters and $\bq,\bp\in\RR^N$ are canonical coordinates and momenta,  was proven in~\cite{PD} to be MS. 
 The kinetic energy $ {\cal T}_\la(\bq,\bp)$ can be interpreted as the one generating the geodesic motion of a particle with unit mass  on a conformally flat space with metric and (non-constant) scalar curvature (\ref{aac})  given by
 \be
 \dd s_\la^2= (1+\la \bq^2)\dd \bq^2 , \qquad  R_\la(\bq)=-\la\,\frac{(N-1)\bigl( 2N+3\la(N-2) \bq^2\bigr)}{(1+\la \bq^2)^3} .
 \label{ad}\nonumber
 \ee
Such a curved space is, in fact,  an $N$D spherically symmetric generalization $\cM_\la$~\cite{plb,annals2009} of the Darboux surface of type III~\cite{Ko72,KKMW03,Pogosyan}.     The   limit $\la\to 0$  of the above expressions  leads to the well known results concerning the (flat) $N$D isotropic harmonic oscillator with frequency $\om$:
   \be
 {\cal H}=\frac 12 \bp^2+\frac 12 \om^2\bq^2 , \qquad  \dd s^2= \dd \bq^2 ,\qquad R=0.
 \label{ae}\nonumber
 \ee
The remarkable point is that    ${\cal H}_\la$ is a MS  Hamiltonian, a fact that can be  stated as follows.

\medskip

 \noindent
{\bf Proposition 1.}~\cite{PD,sigma72011}  {\em  (i) The Hamiltonian ${\cal H}_\la$  (\ref{ac}) is endowed with the following constants of motion ($m=2,\dots,N$):

\noindent
$\bullet$ $(2N-3)$  angular momentum integrals:
\be
  C^{(m)}=\!\! \sum_{1\leq i<j\leq m} \!\!\!\! (q_ip_j-q_jp_i)^2 , \qquad 
 C_{(m)}=\!\!\! \sum_{N-m<i<j\leq N}\!\!\!\!\!\!  (q_ip_j-q_jp_i)^2 ,  \qquad C^{(N)}=C_{(N)} . \label{af}
 \ee
$\bullet$ $N^2$ integrals  which form the ND curved/deformed Demkov--Fradkin tensor ($i,j=1,\dots,N$):
 \be
 I_{\la, ij}=p_ip_j-\bigl(2\la  {\cal H}_\la(\bq,\bp)-\om^2\bigr) q_iq_j , \qquad {\cal H}_\la=\frac 12 \sum_{i=1}^N I_{\la,ii}\, .
\label{ag}\nonumber
\ee
(ii) Each of the three  sets $\{{\cal H}_\la,C^{(m)}\}$,  
$\{{\cal H}_\la,C_{(m)}\}$ ($m=2,\dots,N$) and   $\{I_{\la, ii}\}$ ($i=1,\dots,N$) is  formed by $N$ functionally independent functions  in involution.

\noindent
(iii) The set $\{ {\cal H}_\la,C^{(m)}, C_{(m)},  I_{\la, ii} \}$ for $m=2,\dots,N$ with a fixed index $i$    is  constituted  by $(2N-1)$ functionally independent functions. 
}
\medskip

Let us now consider the   standard definitions for 
the quantum positions $\hbq=(\hq_1,\dots,\hq_N)$,  momenta $\hbp=(\hp_1,\dots,\hp_N)$ and $\nabla=(\frac{\partial}{\partial q_1},\dots, \frac{\partial}{\partial q_N})$ operators ($i,j=1,\dots, N)$:
$$
\hq_i \,\psi(\bq)=q_i\,\psi(\bq),\qquad \hp_i\,\psi(\bq)=-\rmi  \hbar \,\frac{\partial\,\psi(\bq)}{\partial q_i}, \qquad
[\hq_i,\hp_j]=\rmi \hbar\, \delta_{ij},
\qquad \mathbf{q}\cdot\nabla  =\sum_{i=1}^N  q_i\frac\pd{\pd q_i} \, 
.
$$
If  we now apply the conformal Laplacian quantization prescription (\ref{conf}) to ${\cal H}_\la$  (\ref{ac})  we find: 

\medskip
 \noindent
{\bf Proposition 2.}~\cite{annals2011}  {\em   Let   $\hamG$ be the quantum Hamiltonian given by
 \begin{align}
 \hamG &= -\frac{\hbar^2}2\De_{\rm LB,\lambda}+\frac{\omega^2\mathbf{q}^2}{2(1+\lambda\mathbf{q}^2)} -  {\hbar^2\la (N-2)} \left(\frac{  2N+3\la(N-2)\bq^2}{ 8(1+\lambda  {\bq}^2)^3} \right)\, , \nonumber\\
& \mbox{with}\qquad \De_{\rm LB,\lambda}  =   \frac{1}{ (1+\lambda \mathbf{q}^2)}\,\De + \frac{ \lambda(N-2)}{  (1+\lambda\mathbf{q}^2)^2}\, (\mathbf{q}\cdot\nabla ) \, . \label{kn}
 \end{align}
(i) $\hamG$ commutes with the       $(2N-3)$ quantum angular momentum operators $(m=2,\dots,N)$
\be
  \hC^{(m)}=\!\! \sum_{1\leq i<j\leq m} \!\!\!\! ( \hq_i \hp_j- \hq_j \hp_i)^2  , \qquad 
  \hC_{(m)}=\!\!\! \sum_{N-m<i<j\leq N}\!\!\!\!\!\!  ( \hq_i \hp_j- \hq_j \hp_i)^2  ,  \qquad \hC^{(N)}=\hC_{(N)}\, ,
  \label{cb}
 \ee
  as well as with the   $N^2$ curved/deformed Demkov--Fradkin  operators  given by
  \bea
&&  \hI_{{\rm c},\la, ij}= \hp_i\hp_j -(N-2)\frac{\rmi\hbar\la}{2(1+\la\hbq^2)} \,(\hq_i\hp_j +\hq_j\hp_i ) + \frac{(N-2)\hbar^2\la^2\hq_i\hq_j}{(1+\la\hbq^2)^2}\left(1-\frac{N-2}{4} \right) \nonumber   \\ 
 &&\qquad \quad \qquad  - \frac{(N-2)\hbar^2\la}{2(1+\la\hbq^2)}\,\delta_{ij}
   - 2\la  \hq_i\hq_j { \hamG}( \hbq, \hbp)+ \om^2 \hq_i\hq_j ,
\label{kp}\nonumber
\eea
with $ i,j=1,\dots,N$ and such that  ${\hamG}=\frac 12 \sum_{i=1}^N  \hI_{{\rm c},\la,ii}$.  

\noindent
(ii) Each of the three  sets $\{{\hamG},\hC^{(m)}\}$,  
$\{{\hamG},\hC_{(m)}\}$ ($m=2,\dots,N$) and   $\{  \hI_{{\rm c},\la, ii}\}$ ($i=1,\dots,N$) is  formed by $N$ algebraically independent  commuting observables.

\noindent
(iii) The set $\{ { \hamG},\hC^{(m)}, \hC_{(m)},    \hI_{{\rm c},\la, ii}\}$ for $m=2,\dots,N$ with a fixed index $i$    is  formed by $(2N-1)$ algebraically independent observables.

\noindent
(iv) $\hamG$ is formally self-adjoint on the space $L^2(\cM_\la)$,  associated with the underlying Darboux  III space, defined by
\be
\langle \Psi | \Phi \rangle_{\rm c,\la} = \int_{\cM_\la} \overline{{\Psi}(\bq)}\, \Phi(\bq)\, (1+\la\bq^2)^{N/2}\,\dd\bq .
\label{product2}\nonumber
\ee
}
\medskip

The complete solution to the eigenvalue problem along with the corresponding eigenfunctions for the case of {\em positive} deformation parameter $\la$ is summarized in the following statement.

\medskip

 \noindent
{\bf Theorem 3.}~\cite{annals2011}  {\em Let $\hamG$ be the quantum Hamiltonian   (\ref{kn}) with $\la> 0$. Then: 

\noindent
(i) The continuous spectrum of $\hamG$ is given by $[\frac{\om^2}{2\la },\infty)$. There are no embedded eigenvalues and its singular spectrum is empty. 

\noindent
(ii)   $\hamG$ has an infinite number of eigenvalues, all of which are contained in $(0,\frac{\om^2}{2\la })$. Their only accumulation point is $\frac{\om^2}{2\la }$ which is  the bottom of the continuous spectrum. 

\noindent
(iii)  All the discrete  eigenvalues of $\hamG $ are of the form 
\bea
E_{\la,n} \!\!\!&=&\!\!\! - \lambda \hbar^2\left(n + \frac{N}{2}\right)^2 + \hbar \left(n+\frac N 2\right ) \sqrt{\hbar^2 \lambda^2 \left(n+\frac{N}{2}\right)^2+ \om^2 } ,\qquad n\in \NN  \, .
\label{lan} 
\eea
(iv) The   eigenfunction $\Psi_{\LB,\la}$ of $\hamG $ with eigenvalue $E_{\la,n}$  is given by  
\begin{equation}\label{psin}
\Psi_{\LB,\la}(\bq)=(1+\la \bq^2)^{(2-N)/4}\prod_{i=1}^N   \exp\{-\bb^2 q_i^2/2\} H_{n_i}(\bb q_i) ,\qquad  \bb=\sqrt{\frac{\omm}{\hbar}}, \nonumber
\end{equation}
where  $H_{n_i}$  are  Hermite polynomials with   $n_i\in\NN$ such that  $n_1+\cdots+ n_N=n$ and the 
 deformed frequency $\Om$ is defined by
$$
\Om=\sqrt{\om^2-2 \la E_{\lambda,n}} \, .
$$
 }
\medskip

\begin{figure}
\includegraphics[height=8cm,width=11cm]{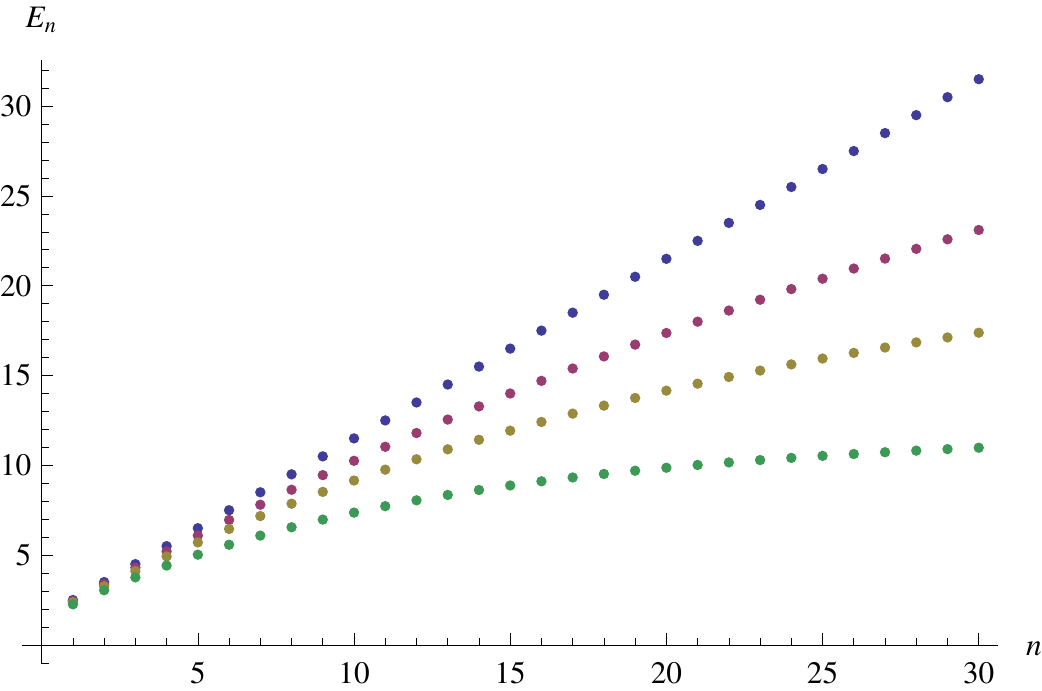}
\caption{The  discrete spectrum  $E_{\la,n}$ (\ref{lan}) for   $0\le n\le 25$, $N=3$,  $\hbar=\om=1$ and $\la=\{0,0.01,0.02,0.04\}$ starting from the upper dot (straight)  line   corresponding to the isotropic harmonic oscillator with $\la = 0$, that is, $E_{0,n}$. In the same order, $E_{\la,0}=\{ 1.5,  1.48, 1.46 , 1.41\}$ and $E_{\la,\infty}=\{ \infty,50,25,12.5\}$.
 \label{figure1}}
\end{figure}

 Moreover,   the bound states of this system satisfy
$$
 E_{\la,\infty}= \lim_{n\to \infty}E_{\la,n}=\frac{\om^2}{2\la}  ,\qquad  \lim_{n\to \infty}(E_{\la,n+1}-E_{\la,n})=0.
 $$
The discrete spectrum  (\ref{lan}) is depicted in  figure~\ref{figure1} as a function of $n$  for several values of $\la$.
 As it was remarked in the introduction, the spectrum turns out to be maximally degenerate since it can be described as a function of just   one quantum number $n\in \NN$.   By taking into account the definition of $n$,  the number of degenerate states
$D(E_{\lambda,n})$ for a given energy level $E_{\lambda,n}$ corresponds to all the possible
combinations of $\{n_i\in\NN\}$ obeying to the constraint $\sum_{i=1}^N n_i = n$, namely
\begin{equation}
D(E_{\lambda,n}) = \frac{(n+N-1)!}{n! (N-1)!} ,
\nonumber
\end{equation}
which for $N=3$ reduces to the well known expression $D(E_{\lambda,n}) =    {(n+2)(n+1)}/{2}$ for the degeneracies of the isotropic oscillator. Therefore, the curved system has a similar set of integrals of the motion as the undeformed one and, as a consequence, it exhibits the same degeneracy.


\section{An exactly solvable deformation of the Coulomb system}

Now we   consider the     $N$D Hamiltonian system given by
\be
\cH_\eta={\cal T}_\eta(\bq,\bp)+{\cal U}_\eta(\bq)=\frac{|\bq|}{2(\eta + |\bq|)}\,\bp^2-\frac{\kk}{\eta + |\bq|} \, ,
\label{otro}
\ee
where $\eta$ and $\kk$ are real parameters. The metric and scalar curvature of the  underlying manifold $\cM_\eta$ turns out to be 
\be
 \dd s_\eta^2=\left(1+  \frac{\eta}{|\bq| }\right) \dd\bq^2,\qquad   R_\eta=\eta (N-1) \,\frac{  4(N-3) r+3\eta(N-2) }{ 4 r  (\eta+  r)^3},\qquad r=|\bq|=\sqrt{\bq^2}\, .
  \label{metr}\nonumber
\ee
We remark that  the system (\ref{otro})  is directly related to a reduction~\cite{IK94} of the geodesic motion on the   Taub--NUT space~\cite{Ma82,AH85,GM86,FH87,GR88,IK95,uwano,BCJ,BCJM,GW07,JL}. In fact,  this system can be regarded as an $\eta$-deformation of the $N$D Euclidean  Coulomb  problem with coupling constant $\kk$, since the  limit $\eta\to0$ yields
$$
{\cal H} =\frac12\, { \bp^2} -\frac{\kk}{ |\bq|}  , \qquad  \dd s^2= \dd \bq^2 ,\qquad R=0. 
$$

Remarkably enough, the Hamiltonian $\cH_\eta$
turns out to be a    MS  classical system,  and this result can be summarized as follows. 

\medskip

 \noindent
{\bf Proposition 4.}~\cite{sigma72011}  {\em   (i) The Hamiltonian  $
\cH_\eta$  (\ref{otro})     is endowed with   the    $(2N-3)$   angular momentum integrals   (\ref{af})  and Poisson-commutes with 
 the $\cR_{ \eta, i}$ components ($i=1,\dots,N$) of the    Runge--Lenz $N$-vector  given by
$$
\cR_{\eta, i}=\sum_{j=1}^N p_j ( q_j p_i - q_i p_j )+\frac{ q_i}{|\bq|} \left(\eta \cH_\eta(\bq,\bp)+\kk\right).
$$
 (ii) The set $\{ {\cal H}_
\eta,C^{(m)}, C_{(m)},  \cR_{ \eta . {i}} \}$     with $m=2,\dots,N$ and a fixed index $i$    is formed by    $(2N-1)$ functionally independent functions. 
}
  
  \medskip

 We also recall that the classical  system   $\cH_\eta$ has been fully solved in~\cite{ragnisco}.
Next  the quantum   counterpart  of (\ref{otro}) can be obtained by applying (\ref{conf}), and reads: 
  \medskip

 \noindent
{\bf Proposition 5.}~\cite{annals2014}  {\em (i) The quantum Hamiltonian   $\hamGb$   given by
\bea
&&\hamGb= -\frac{\hbar^2}2\De_{\rm LB,\eta} -\frac{k }{\eta + |\bq|} +\hbar^2 \eta (N-2)\,\frac{  4(N-3)|\bq|+3\eta(N-2)  }{32|\bq|(\eta+|\bq|)^3} \, ,
\nonumber\\[2pt]
&&\mbox{with} \qquad \Delta_{\rm LB,\eta}= \frac{ |\bq|}{\eta + |\bq|} \, \De - \frac{\eta(N-2)}{2|\bq| (|\bq|+\eta)^2}\, (\mathbf{q}\cdot\nabla ) \, ,
\label{kn2}
\eea
  commutes with the $(2N-3)$  quantum angular momentum operators  (\ref{cb})     as well as with 
  the following $N$ Runge--Lenz operators $(i=1,\dots,N)$:
  \bea
 && {  \normalsize{ \hat{\cal R}_{{\rm c},\eta, i}= \frac{1}{2}\sum_{j=1}^N\left(  \hp_j + \rmi  \hbar \, \eta\, \frac{(N-2)\hq_j}{4
(\eta+|\hbq|)\, \hbq^2}  \right) ( \hq_j \hp_i - \hq_i \hp_j ) }}\nonumber\\[2pt]
&&\qquad\qquad  + \frac{1}{2} \sum_{j=1}^N  ( \hq_j \hp_i - \hq_i \hp_j ) \left(  \hp_j + \rmi  \hbar \, \eta\, \frac{(N-2)\hq_j}{4
(\eta+|\hbq|)\,\hbq^2}  \right)   +\frac{ \hq_i}{|\hbq|} \left(\eta \hH_{\rm c,\eta}(\bq,\bp)+\kk\right).
\nonumber
\eea

\noindent
 (ii) Each of the three  sets $\{{\hamGb},\hC^{(m)}\}$,  
$\{{\hamGb},\hC_{(m)}\}$ ($m=2,\dots,N$) and   $\{  \hat{\cal R}_{{\rm c},\eta,i}\}$ ($i=1,\dots,N$) is  formed by $N$ algebraically independent  commuting operators.

\noindent
 (iii) The set $\{ { \hamGb},\hC^{(m)}, \hC_{(m)},   \hat{\cal R}_{{\rm c},\eta,i}\}$ for $m=2,\dots,N$ with a fixed index $i$    is  formed by $(2N-1)$ algebraically independent operators.

\noindent
 (iv) $\hamGb$ is formally self-adjoint on the Hilbert space $L^2(\cM_\eta)$
with the scalar product 
\be
\langle \Psi | \Phi \rangle_{\rm c,\eta} = \int_{\cM_\eta} \overline{{\Psi}(\bq)}\, \Phi(\bq)\, 
\left(1+\frac \eta{|\bq|} \right)^{N/2}\,\dd\bq .
\nonumber
\ee
}

For a {\em positive} value of the deformation parameter $\eta$, the  complete solution of the eigenvalue problem for this quantum mechanical deformed Coulomb problem is the following.

  \medskip

 \noindent
{\bf Theorem 6.}~\cite{annals2014}  {\em Let $\hamGb$ be the quantum Hamiltonian (\ref{kn2})  with $k>0$ and $\eta>0$. Then: 

\noindent
  (i) The continuous spectrum  of $\hamGb$ is given by $[0,\infty)$. There are no embedded eigenvalues and the singular spectrum is empty. 

\noindent
 (ii) $\hamGb$ has an infinite number of eigenvalues $E_{\eta,n, l}$, depending only
on the sum $(n+l)$ and accumulating at $0$.

\noindent
  (iii)  The eigenvalues $E_{\eta,n,l}$ of $\hamGb$ are of the form  
\begin{equation}
\label{eq13}
E_{\eta,n,l} =\frac{-k^2}{\hbar^2\left(n+l+\frac{N-1}{2} \right)^2+k \eta + \sqrt{\hbar^4\left(n+l+\frac{N-1}{2}\right)^4+2\,\hbar^2k \eta \left(n+l+\frac{N-1}{2}\right)^2}} \, ,
\end{equation}
such that the   radial  eigenfunction $\Phi_{{\rm c}, \eta}(r) $ of $\hamGb$   with eigenvalue $E_{\eta,n,l}$ reads
$$
\Phi_{{\rm c, \eta}}(r) =\left(1+\frac{\eta}{r} \right)^{\frac{2-N}{4}} r^l \exp\left( - \frac{K r}{ \hbar^2\left(n+l+\frac{N-1}{2}\right)}  \right)   \,L_n^{2l+N-2}\left( \frac{2 K r}{\hbar^2\left(n+l+\frac{N-1}{2}\right)}\right) ,
$$
where   $L_n^\alpha$ are generalized Laguerre polynomials and the deformed coupling constant  $K$ reads
$$
K = k + \eta\, E_{\eta,n,l} .
 $$
}

Since   $\hamGb$ is a Hamiltonian with radial symmetry,   its complete eigenfunction   is so given by $\Psi_{\LB,\eta}= \Phi_{{\rm c} ,\eta}(r)Y_{\boldsymbol{l}}(\boldsymbol{\te})$ where $Y_{\boldsymbol{l}}(\boldsymbol{\te})$ denotes the usual hyperspherical harmonics, $\boldsymbol{\te}=(\te_1,\dots,\te_{N-1})$ and $\boldsymbol{l}$ is a vector of $N-1$ quantum numbers $\boldsymbol{l} = (l_2, \dots, l_{N-1}, l_{N}=l)$ such that (see~(\ref{cb}))
\begin{equation}
{\hat C}^{(m)}     Y_{\boldsymbol{l}}(\boldsymbol{\theta}) = \hbar^2 l_m (l_m + m-2)
Y_{\boldsymbol{l}}(\boldsymbol{\theta}), \qquad l_m-l_{m-1} \geq 0 ,\qquad m=2,\dots N. \nonumber
\end{equation}


\begin{figure}
\begin{center}
\includegraphics[height=8cm]{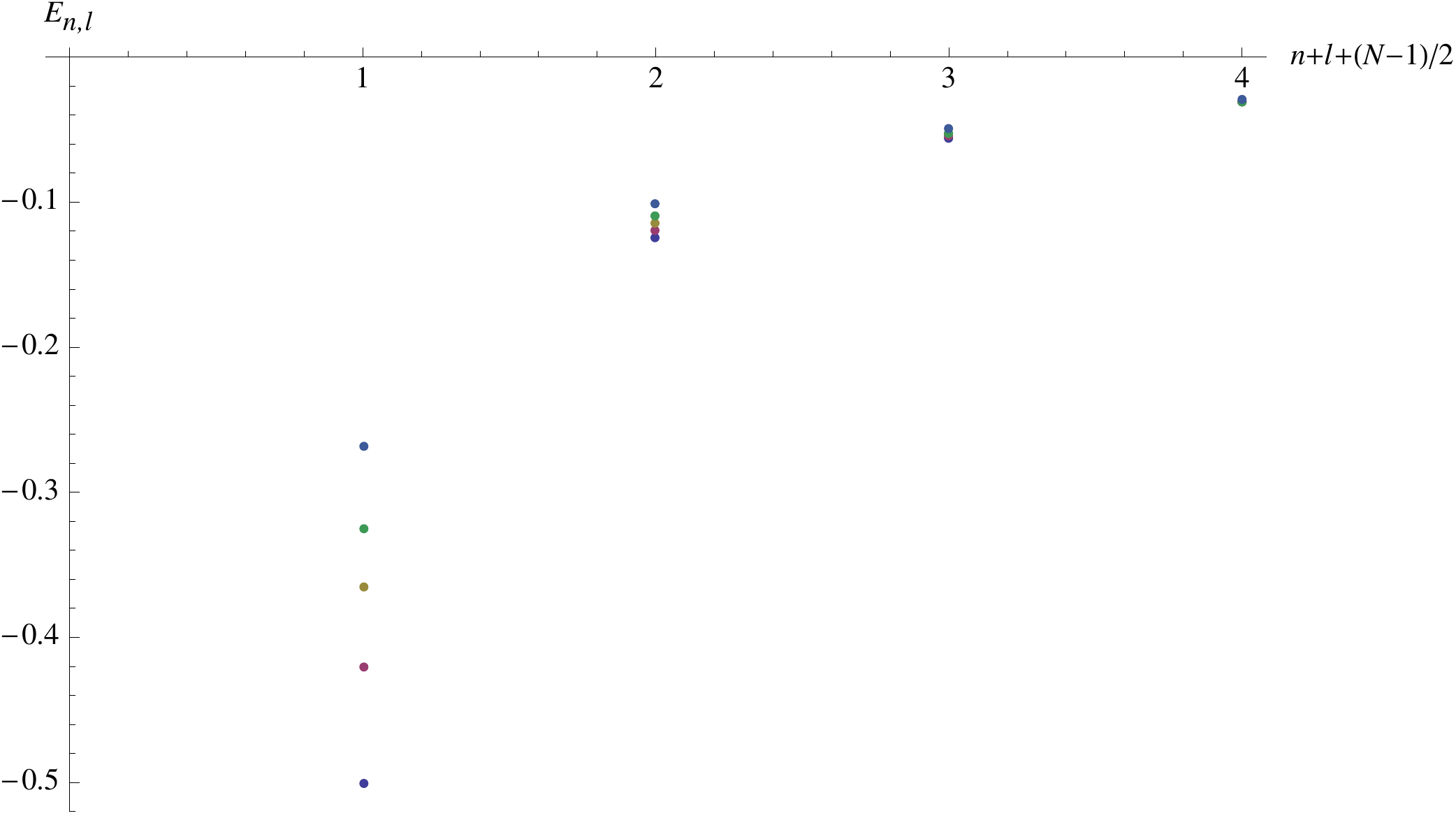}
\caption{Discrete spectrum (\ref{eq13})  for the fundamental and the three first excited states of the Hamiltonian $\hamGb$ (\ref{kn2}) when $\eta=\{0, 0.2, 0.4, 0.6, 1\}$ with $\hbar=k=1$ and $N\geq 3$. Note that the effect of the $\eta$ deformation is quite strong for the fundamental state, since it comes from the shift $r\to r+\eta$ in the usual Coulomb potential. 
 \label{figure2}}
\end{center}
\end{figure}


Notice also that the bound states of this system satisfy
\be
  \lim_{n,l\to \infty}E_{\eta,n,l}= 0 ,\qquad  \lim_{\mathfrak{n}\to \infty}(E_{\eta,\mathfrak{n}+1}-E_{\eta, \mathfrak{n}})=0, \qquad \mathfrak{n} =n+ l.
  \nonumber
\ee
As expected, the limit $\eta\to
0$ of $E_{\eta,n,l}$ provides  the well known formula for the standard
Coulomb eigenvalues $ E_{0,n,l}$
$$
E_{0,n,l}=-\frac{k^2}{ 2 \hbar^2 \left (n+l+\frac{N-1}{2} \right)^2}\,.
$$
And we find that the  perturbative series for the eigenvalues of the deformed system $\hamGb$  (\ref{kn2})  reads
\begin{equation}
\nonumber
E_{\eta, n,l} =E_{0,n,l}  + \eta\,\frac{ k^3}{2 \hbar^4 \left(n+l+\frac{N-1}{2}\right)^4} 
-\eta^2\, \frac{5 k^4}{8 \hbar^6 \left(n+l+\frac{N-1}{2}\right)^6} + O(\eta^3) .
\end{equation}
In figure 2 the eigenvalues of the fundamental and of the first three excited states are plotted for different values of the deformation parameter $\eta$.

 As we can see from (\ref{eq13}) the spectrum is maximally degenerate as, again, it depends on a unique principal quantum number $\mathfrak{n} =n+l$. The degeneracy $D(E_{\eta, \mathfrak{n}})$ of a given energy level $E_{\eta, \mathfrak{n}}$ can be computed straightforwardly by taking into account  that the cardinality $D(L_l)$ given by the set of the hyperspherical harmonics $\{ Y_{\boldsymbol{l}}(\boldsymbol{\theta}) \}$ having the same quantum number $l$ and  such that $\hat{C}^{(N )}\{ Y_{\boldsymbol{l}}(\boldsymbol{\theta}) \} =\hbar^2  l (l+N-2) \{ Y_{\boldsymbol{l}}(\boldsymbol{\theta}) \}$ reads~\cite{sphericalharmonic} 
\begin{equation}
D(L_l) = \frac{(2l+N-2) (l+N-3)!}{l! (N-2)!} \,.\nonumber
\end{equation}
From it we obtain that
\begin{equation}
D(E_{\eta,\mathfrak{n}} )= \sum_{l=0}^{\mathfrak{n}} D(L_l )= \frac{(2
\mathfrak{n} +N - 1) (\mathfrak{n} + N - 2)!}{\mathfrak{n}! (N-1)!}.\nonumber
\end{equation}
In particular, for $N=3$ we obtain $
D(E_{\eta,\mathfrak{n}} )=   (\mathfrak{n} + 1)^2
$, which coincides with the degeneracy of the energy levels of the undeformed Coulomb problem.


\section{Generalization}

So far we have reviewed  some {\em specific} exactly solvable deformations of the oscillator and Coulomb potentials, which can be regarded as 
 the most natural MS deformations beyond constant curvature. Nevertheless, there are more possible generalizations within this framework that 
 preserves the classical MS property and that would lead to other   exactly solvable deformed  oscillator and Coulomb systems. These arise within the classification of Bertrand Hamiltonians formerly introduced in~\cite{Perlick} and further developed in~\cite{Bertrand,danilo, danilo2}. Such systems are MS and their underlying 
 Bertrand spaces are spherically symmetric ones.   If we require to keep {\em quadratic} integrals of motion, so generalizing the Demkov--Fradkin tensor and the Runge--Lenz $N$-vector, it can be shown   that there   only exists {\em one} possible generalization of  the deformations of the oscillator and Coulomb systems here studied that depends on {\em two} deformation parameters.
 
 In particular, the two-parameter MS deformation of the oscillator system turns out to be
  \be
{\cal H}_{\la,\xi}(\bq,\bp)={\cal T}_{\la,\xi}(\bq,\bp)+{\cal U}_{\la,\xi}(\bq)=
\frac{(1-\xi \bq^4)^2\bp^2}{2(1+\la \bq^2+\xi \bq^4 )}+\frac{ \om^2 \bq^2}{2(1+\la \bq^2+\xi \bq^4 )}  ,
\nonumber
\ee
where $\xi$ is a real parameter. Obviously, the limit $\xi\to 0$ gives rise to the Hamiltonian ${\cal H}_{\la}$  (\ref{ac}) 
 The  underlying manifold  $\cM_{\la,\xi}$  is endowed with a conformally flat metric given by
 $$
  \dd s_{\la,\xi}^2=\frac{(1+\la \bq^2+\xi \bq^4)}{(1-\xi \bq^4)^2}   \,\dd \bq^2 .
 $$
And the corresponding scalar curvature (\ref{aac}) reads
\bea
&&R_{\la,\xi}(r)=-\frac{(N-1)}{(1+\la r^2+\xi r^4)^3}\bigg\{ N\big(2+3\la r^2+ 6 \xi r^4+\la \xi r^6 \big) \big( \la + 3 \la \xi r^4 + 2 \xi r^2 (3+\xi r^4)\big)   
\nonumber\\[2pt]
&& \qquad\qquad\qquad\qquad\qquad\qquad    -6r^2(\la^2- 4 \xi)(1-\xi r^4)^2 \bigg\} ,
\nonumber
\eea
where recall that $r=|\bq|=\sqrt{\bq^2}$.

As far as the Coulomb system is concerned, the resulting  two-parameter MS deformation  is given by
$$
\cH_{\eta,\zeta}={\cal T}_{\eta,\zeta}(\bq,\bp)+{\cal U}_{\eta,\zeta}(\bq)=\frac{(1-\zeta \bq^2)^2|\bq|}{2(\eta + |\bq| + \eta\zeta \bq^2) }\,\bp^2-\frac{\kk(1+\zeta \bq^2)}{(\eta + |\bq| +\eta\zeta\bq^2)} \, ,
$$
which generalizes the one-parameter Hamiltonian ${\cal H}_{\eta}$ (\ref{otro}).
Hence the metric of the underlying spherically symmetric space  $\cM_{\eta,\zeta}$ and its scalar curvature are found to be 
$$
 \dd s_{ \eta,\zeta}^2=\frac{(\eta + |\bq| + \eta\zeta \bq^2)}{(1-\zeta \bq^2)^2|\bq|}\, \dd \bq^2  \, ,
 $$
\bea
&&R_{ \eta,\zeta}(r)=-\frac{(N-1)}{ 4 r(\eta + r + \eta \zeta r^2 )^3}\bigg\{ 6 \eta (1-\zeta r^2)^2\bigg(\eta + r\big(2 +\zeta r(6 \eta + r [ 2+\eta \zeta r] ) \big)\bigg )
\nonumber\\[2pt]
&& \qquad\qquad\qquad  -N\big(3\eta + r (4+ \eta\zeta r [6-\zeta r^2 ] )\big) \big(\eta -\zeta r^2 (6\eta + r [4 + 3\eta\zeta r])  \big)
 \bigg\} .
\nonumber
\eea
 It is worth stressing that  $\cM_{\eta,\zeta}$ turns out to be the  $N$D spherically symmetric generalization   of the Darboux surface of type IV~\cite{Ko72,KKMW03,Pogosyan} constructed in~\cite{plb, annals2009}.
 
 Consequently, by applying the conformal    Laplacian quantization (\ref{conf}) to the above two-parameter Hamiltonians,  new exactly solvable systems, $\hat\cH_{\rm c,\la,\xi}$ and $\hat\cH_{\rm c,\eta,\zeta}$, would be obtained as deformations of   the oscillator and Coulomb systems. Their solution would generalize the results presented in theorems 3 and 6. Work on this line is currently in progress.


\section*{Acknowledgments}

This work was partially supported by the Spanish MINECO through the Ram\'on y Cajal program (A.E.) and under grants MTM2013-43820-P (A.B and F.J.H.) and   FIS2011-22566 (A.E.), by the Spanish Junta de Castilla y Le\'on under grant  BU278U14 (A.B., A.E. and F.J.H.), 
 by  the ICMAT Severo Ochoa under grant SEV-2011-0087 (A.E.),   and by  a postdoctoral fellowship  from the Laboratory of
Mathematical Physics of the CRM, Universit\'e de Montr\'eal (D.R.).


\end{document}